# COLLECTIVE ACHIEVEMENT OF MAKING IN COSPLAY CULTURE

Rie Matsuura, Daisuke Okabe

Tokyo City University
Yokohama, Kanagawa, 224-8551, JAPAN
g1383117@tcu.ac.jp, okabe@tcu.ac.jp

## INTRODUCTION

Cosplay is a Japlish combining of Costume and Play. They are aimed an expression of affection for anime and manga's story and characters. The majority of cosplayers in Japan are women, mostly high school and college students and people in their twenties.

Cosplay is a female DIY culture. The DIY spirit embodied by this practice has become a standard in the cosplay community. Cosplay can be seen at dedicated cosplay parties at amusement parks and events for fanzine sale. They gather in cosplay events with costumes of their own sawing or ready-made ones, and are photographed by the audiences or each others. Cosplay events and dedicated SNSs for cosplayer's are a valuable venue for exchanging information, peer reviewing, and collective learning from each other about costume making, as well as for evaluating each other's work.

Ito et al (2013) study the possibilities of learning with friends and peers in fan culture in the US. They said that learning in fan culture is characterized as peer-based and friendship-based learning. First, according to their perspectives, we will indicate the cosplay community characterized as interest-driven, peer-based reciprocal learning environment. For example, some schools are engaged in efforts to design a learning environment that encourages peer-based, reciprocal learning (for example, Johnson et al. 1993). This kind of learning environment goes against the dominance of top-down instruction that has been institutionalized in most schools.

Then, we will focus on how do women's cosplayers socialize and learn with their friends and in information and knowledge ecology from the point of view of Brunner's "Scaffolding" theory. The original notion of Scaffolding assumed that more the knowledgeable person, such as a teacher, helps individual learners, providing them with the support they need to move forward (e.g., Bruner, 1975; Wood et al., 1976). We think that the notion of Scaffolding might be expanded in this networked and connected world.

Compared to learning environments in most schools, the cosplay community has always been based on peer-based scaffolding environment, with members creating their own skills and levels of conduct. As a purpose of this study, we might look to them as models for designing interest-driven communities and collaborative learning environments, and we'd like to expand "Scaffolding" concept to more reciprocal one, rather than an interaction between individual learner and a knowledgeable person.

## RESEARCH

The data introduced in this section were collected using ethnographic methods, a combination of interviews and field observations. Ten informants, all female, were interviewed (see Table 1). Because the majority of cosplayers in Japan are female, we focused our research on female cosplayers. The study was conducted between August 2011 and December 2013. Informants included students and employees. They were restricted to cosplayers in their twenties since most active cosplayers are teens and young adults. Info.1 was our initial contact, and we used snowball sampling to recruit the other informants and to maximize rapport during interviews and field observations. During field observations at cosplay events, we used digital cameras, camcorders, digital voice recorders, and field notes to record the flow of events; however, as camcorders are often forbidden at cosplay events there were times we were unable to make recordings. After-event parties are also common, so we recorded conversations that took place at after-event parties when possible. The interviews focused on the informant's experiences from the time she first became a participating member in the cosplay community up to the time of the interview. Although we prepared a set of questions to be asked in sequence to follow a rough time line based on the informant's experience, we focused on facilitating the informant's narrative and did not correct digressions when and if they occurred. Representative topics in the interview are as follows:
1) Perceived changes in herself as she gained online/offline cosplay experience
2) Opinions regarding what constitutes taboo behavior at cosplay events

3) Characteristics of a cosplayer who is popular within the SNS community and so on

*Table 1: The list of informants*

| ID | Age | Sex | Occupation | Years of experience |
|---|---|---|---|---|
| info. 1 | 22 | Female | College student | 4 |
| info. 2 | 22 | Female | College student | 4 |
| info. 3 | 22 | Female | Undergraduate | 6 |
| info. 4 | 25 | Female | Corporate employee | 6 |
| info. 5 | 21 | Female | Professional school student | 3 |
| info. 6 | 22 | Female | College student | 8 |
| info. 7 | 22 | Female | College student | 6 |
| info. 8 | 20 | Female | College student | 1 |
| info. 9 | 23 | Female | Undergraduate | 8 |
| info.10 | 20 | Female | College student | 8 |

**RESULT**

The interview transcripts were analyzed according to the "Steps for coding and Theorization(SCAT)" method, a qualitative data analysis technique by Otani (2008). It consists of steps of coding from open to selective, like grounded theory approach. Starting from the interview data, we picked out some words from informants' utterances to generate some concepts. We got 9 categories through SCAT analysis as below: 1) DIY Ethics, 2)Shared Rules and Codes of Conduct, 3) Peer Review, 4)Rejecting Commercial and Mainstream Cosplay, 5) Reciprocal Learning, 6) Sharing Cosplay Knowledge and Information on SNSs, 7) Learning from Others' Digital Data, 8)SNSs as Scaffolding System, 9)Standardization and Creating New Tasks.

In this paper, we'd like to introduce an interview data related to three categories about 6), 8) and 9). Because dedicated SNSs for cosplayers are a valuable venue for exchanging information and collective learning about costume making.

We could see the following utterance at the interview.

*interview data 1: "You can see how to make the weapon on Twitter. So it's possible to make it, but most people don't pay attention to the painting. If you don't paint it properly, it doesn't match the character. When I make this weapon. I'm really careful with the painting process." (Info.10)*

You can observe these reciprocal situations on online. "Cure" is a community SNS site for cosplayers. They upload photos. Cosplayers use Twitter, and upload photos about the making of costumes and items. Cosplayers use other cosplayers' photos on the web as a reference. They expect that other cosplayers will see their photos as a reference in turn (see interview data 1 and figure 1). Cosplayers use pictures on the Web as "scaffolding" and they interact with each other, and they achieve "what they can not achieve by themselves". For example, when a famous cosplayer uploads photos of how to make difficult costumes, those pictures are retweeted instantly. Then, this knowledge is shared among the other cosplayers and they can challenge to make more difficult costumes and orient to new tasks for future cosplay activities. We consider the whole system or whole activities including this kind of interaction to be "Scaffolding".

**CONCLUSION**

In traditional scaffolding, a supporter supports an explicit learner the concept of scaffolding includes direct interaction with learner. On the other hand, cosplayers use SNS sites for archiving pictures as scaffolding, using another cosplayers photo as a reference. They can achieve "what they can not achieve by themselves." We can expand the concept of scaffolding as a whole system including both SNS and peer interaction with cosplayers.

The cosplay community, in contrast, has always been based on peer-based, reciprocal learning, with members creating their own rules and codes of conduct. School learning, with its superior/subordinate teacher/learner relationships, has become the norm, and interest-driven learning has become marginalized. That may be why cosplayers appear so odd to the main-stream. Rather than marginalize or stigmatized these groups, however, we might look to them as models for designing interest-driven communities and collaborative learning environments.